\documentclass[twocolumn,floatfix, nofootinbib,prd,preprintnumbers,showpacs,showkeys,superscriptaddress,longbibliography]{revtex4-2}

\usepackage{amssymb}
\usepackage[mathscr]{euscript}
\usepackage{amsmath}
\usepackage{graphicx}
\usepackage{dcolumn}
\usepackage{bm}
\usepackage{epsfig}
\usepackage{latexsym,mathrsfs}
\usepackage{graphicx}
\usepackage{color}
\usepackage{hyperref}
\usepackage{float}
\usepackage[thinlines]{easytable}
\usepackage{multirow}
\usepackage{diagbox}
\usepackage{varwidth}
\usepackage{tikz}
\usepackage{tikz-cd}

\newcolumntype{M}[1]{>{\centering\arraybackslash}m{#1}}
\newcolumntype{N}{@{}m{0pt}@{}}

\hypersetup{
    colorlinks=true,
    linkcolor=red,
    citecolor=blue,
}

\usepackage{physics}
\usepackage{csquotes}
\usepackage{mathtools}
\usepackage[caption=false]{subfig}
\usepackage{hyperref}
\usepackage{url}
\usepackage{xcolor}
\usepackage{color}
\usepackage{soul}
\definecolor{amaranth}{rgb}{0.9, 0.17, 0.31}
\definecolor{purple(munsell)}{rgb}{0.62, 0.0, 0.77}
\definecolor{americanrose}{rgb}{1.0, 0.01, 0.24}
\definecolor{palatinateblue}{rgb}{0.15, 0.23, 0.89}
\definecolor{royalblue(web)}{rgb}{0.25, 0.41, 0.88}
\definecolor{hanpurple}{rgb}{0.32, 0.09, 0.98}
\definecolor{beaublue}{rgb}{0.74, 0.83, 0.9}
\definecolor{carminered}{rgb}{1.0, 0.0, 0.22}
\definecolor{brightpink}{rgb}{1.0, 0.0, 0.5}
\definecolor{vividviolet}{rgb}{0.62, 0.0, 1.0}

\definecolor{electron}{rgb}{1.0, 0.67, 0.22}
\hypersetup{ linktoc=all,
    colorlinks, linkcolor={palatinateblue},
    citecolor={brightpink}, urlcolor={amaranth}}

\newcommand{\calW}{\mathcal{W}}

\newcommand{\Z}{\mathbb{Z}}
\newcommand{\R}{\mathbb{R}}

\newcommand{\e}{\mathrm{e}}
\renewcommand{\i}{\mathrm{i}}
\newcommand{\beq}{\begin{equation}}
\newcommand{\eeq}{\end{equation}}

\begin{document}

\preprint{FTPI-MINN-25-01}
\preprint{UMN-TH-4416/24}

\title{Degenerate Domain Walls in Supersymmetric Theories}


\author{Shi Chen}
\email{chen8743@umn.edu}
\affiliation{Department of Physics, School of Physics and Astronomy,\\
University of Minnesota, Minneapolis, MN 55455, USA}

\author{Evgenii Ievlev}
\email{ievle001@umn.edu}
\affiliation{Department of Physics, School of Physics and Astronomy,\\
University of Minnesota, Minneapolis, MN 55455, USA}
\affiliation{William I. Fine Theoretical Physics Institute, School of Physics and Astronomy,\\
University of Minnesota, Minneapolis, MN 55455, USA}

\author{Mikhail Shifman}
\email{shifman@umn.edu}
\affiliation{Department of Physics, School of Physics and Astronomy,\\
University of Minnesota, Minneapolis, MN 55455, USA}
\affiliation{William I. Fine Theoretical Physics Institute, School of Physics and Astronomy,\\
University of Minnesota, Minneapolis, MN 55455, USA}


\begin{abstract}
In supersymmetric Yang-Mills theories (SYM) tension-degenerate domain walls are typical. Adding matter fields in fundamental representation we arrive at supersymmetric QCD (SQCD) supporting similar walls. We demonstrate that the degenerate domain walls can belong to one of two classes: (i) locally distinguishable, i.e. those which differ from each other locally (which could be detected in local measurements); and (ii) those which have identical local structure and are differentiated only topologically, through a judicially chosen compactification of $\mathbb{R}^4$. Depending on the number of flavors $F$ and the pattern of Higgsing both classes can coexists among SQCD $k$ walls interpolating between the vacua $n$ and $n+k$. We prove that the overall multiplicity of the domain walls obtained after accounting for both classes is $\nu_{N,k}^\text{walls}= N!/\big[(N-k)!k!\big]$, as was discovered previously in limiting cases. (Here $N$ is the number of colors.) Thus, $\nu_{N,k}^\text{walls}$ is a peculiar index. For the locally distinguishable degenerate domain walls we observe two-wall junctions, a phenomenon specific for supersymmetry with central extensions. This phenomenon does not exist for topological replicas.
\end{abstract}

\maketitle

\section{Introduction}

Supersymmetric  
field theories have miraculous properties -- this  became clear from the very inception. For instance, the foundational algebra of supersymmetry \cite{Golfand:1971iw,Wess:1974tw}
\begin{equation}
\{Q_\alpha\,\bar{Q}_{\dot\beta}\} = 2 P_{\alpha\dot\beta}
\label{one}
\end{equation}
implies that the vacuum energy exactly vanishes. (Here $Q,\,\bar{Q}$ are conserved supercharges and $P$ is the energy-momentum operator.)
Multiple discrete vacua (ground states) are typical for many supersymmetric theories; they are degenerate and, therefore, domain walls interpolating between them must exist. If there are several {\em distinct} domain walls interpolating between one and the same pair of vacua their tensions may be degenerate too
provided the algebra \eqref{one} admits extensions by central (brane) charges, for instance,  for the domain walls they take the form 
\begin{equation}
 \{Q_\alpha\,{Q}_{\beta}\} = Z_{\alpha\beta}
 \label{two}
 \end{equation}
where $Z_{\alpha\beta}$ is a set of numbers depending only on the boundary conditions. Under certain circumstances the domain wall tensions
equal $Z$ -- this phenomenon is referred to as the Bogomol'nyi-Prasad-Sommerfeld (BPS) saturation \cite{Bogomolny:1975de,Prasad:1975kr}. The BPS saturated walls are degenerate. 

In this paper after a brief review of supersymmetric domain walls we will focus on the multiplicity of the degenerate BPS walls in supersymmetric Yang-Mills theories (SYM).

\begin{figure}[ht]
    \centering
    \includegraphics[width=0.3\textwidth]{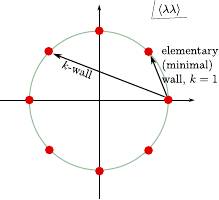}
    \caption{\small
    	Distinct vacua (red closed circles) in the complex plane of $\expval{\lambda\lambda}$ in SYM ($\langle\lambda\lambda\rangle$ is the gluino condensate).
        Arrows indicate the vacua between which the domain walls interpolate.
    	}
    \label{lambdalambda}
\end{figure}

According to Witten's index \cite{Witten:1982df},  SYM theory with the gauge group SU$(N)$ has $N$ distinct discrete vacua labeled by the value of the gluino condensate $$\langle\lambda\lambda\rangle_k = \langle\lambda\lambda\rangle_0 \exp{\frac{2 \pi i k}{N}}\, \,\,\,\, k = 0, 1,2, ... N-1 ,$$ see Fig. 
\ref{lambdalambda}. The domain walls interpolating between the vacua $n$ and $n+k$ are called the $k$ walls. The minimal tension walls
correspond to $k=1$ and are called elementary.

Shortly after the discovery of the brane charge $Z_{\alpha\beta}$ in SYM (it is absent at the classical level and emerges as a quantum anomaly \cite{Dvali:1996xe}) the BPS domain walls were found \cite{Kovner:1997ca}. Later Witten identified them as branes \cite{Witten:1997ep}.
In 2001 Acharya and Vafa demonstrated  (from a wrapped $D$-brane construction with a  toroidal compactification of $\mathbb{R}^4$) \cite{Acharya:2001dz}  that in pure SYM the domain wall world-volume theory is topological, namely, 
 U$(k)$  Chern-Simons   at level $N$, (usually referred to as ${\rm CS}_{k,\,N}$)
which gives rise to $C^N_k$ vacua, where $C^N_k$ is Newton's combinatorial factor.
In this way, they calculated the  multiplicity of degenerate walls in SYM,
 \begin{equation}
\nu_{N,k}^\text{walls}= \frac{N!}{k!(N-k)!} \,.
\label{three}
 \end{equation}
  Here and below $\nu$ stands for multiplicity while the subscripts indicate the gauge group and the wall type.
We emphasize that the Acharya-Vafa walls
are  {\em locally} indistinguishable; i.e. are the same in local measurements and  differ only globally.

Somewhat earlier another strategy  was developed in \cite{Kovner:1997ca}.
It was suggested to deform SYM by adding matter fields in such a way that the gauge sector becomes fully Higgsed, allowing one to work in the weak coupling regime.
For SU(2)$_{\rm gauge}$  two distinct elementary walls with non-identical  internal structure were identified  (in this case only elementary walls exist). 
The study was extended to  SU$(N)$ SQCD by Ritz et al \cite{Ritz:2002fm}.
At the  focus of this paper was  the matter sector consisting of $F$ flavors with $F=N$. 
The matter mass term was assumed to be small, $m\ll\Lambda$ ($\Lambda$ is the dynamical scale of SQCD\footnote{Note that, generally speaking, the definition of the dynamical scale $\Lambda$ depends on the number of colors $N$ and the number of fundamental flavors $F$ as 
$\Lambda^{3N-F} = M_\text{uv}^{3N-F} \dfrac{1}{g^{2N}(M_\text{uv})} e^{ 2 \pi i \tau (M_\text{uv}) }$,
where $M_\text{uv}$ is the UV cutoff and $\tau = \frac{4 \pi i}{ g^2 } + \frac{\theta}{2\pi}$ is the complexified gauge coupling.
The values of $N$ and $F$ in each particular case here are self-evident from the context.
}.)
which guarantees weak coupling.

The full quantum flavor symmetry is manifest, and the pattern of Higgsing of {\em all} gauge bosons is straightforward.  It was  found that wall solutions break the flavor symmetries in such a way that the world volume theory, after factoring out the translational mode, is an ${\cal N} = 1$ Grassmannian sigma model in 2+1D. The vacuum states of this theory therefore count the number of BPS wall supermultiplets.
The Witten index in the latter determines the multiplicity of the degenerate walls. In particular,  $\mathbb{CP}(N-1)$ model emerges on the world volume of the {\em elementary} walls -- it has $N$ vacua. In the generic Grassmannian model the Witten index and, hence, the wall multiplicity,  coincide with \eqref{three}. 
The $\nu_{N,k}$ degenerate domain walls obtained in this way are {\em locally distinguishable}.

\begin{figure}[t]
 \centering
    \includegraphics[width=0.3\textwidth]{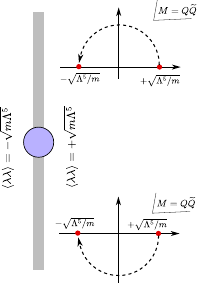}
    \caption{\small 
    	Junction of two domain walls in SU(2) SQCD with one flavor.
    	The gluino condensate $\langle\lambda\lambda\rangle$ is the order parameter marking two distinct vacua. The gray line separating two vacua represents the domain walls. The blue closed circle is the two-wall junction. The walls above and below the junction are different.
    	On the right  two wall trajectories  are presented which show the difference between the walls. $M$ is a gauge invariant  order parameter bilinear in the matter fields.
    	}
    \label{2w-junc}
\end{figure}

The transition domain between the weak coupling SQCD and strong coupling SYM in the limit $m\to\infty$ as well as a related issue of two-wall junctions remained terra incognita for two decades. Locally distinguishable exactly degenerate domain walls can form two-wall junctions\footnote{Such junctions cannot exist in non-supersymmetric theories which usually do not provide exact degeneracy of the wall tensions. } of the type shown in Fig. \ref{2w-junc}. 
At the same time in SYM such junctions are not supported because the degenerate walls all {\em locally identical}. 

Is it possible to reconcile these two facts?
If yes, how?

In this paper we address both topics and develop a general picture of the phenomenon. We argue that both methods of counting -- through pure SYM with CS on the world volume on one hand and added matter sectors on the other, produce the same answer quoted in \eqref{three}
although physics is different. One of our main results is summarized in \eqref{domain_wall_theory} presenting  the wall world volume effective theory in SQCD on $\mathbb{R}^4$ for the arbitrary value of 
$F\leqslant N$. 
For an extended and more technical version see Ref.~\cite{Chen:2024jyn}.

The counting in SYM is  topological, and formula \eqref{three} gives the topological multiplicity in the bulk theory compactified on a torus.
Adding matter fields makes degenerate domain walls locally distinguishable which manifests itself in two-wall junctions. They exist! This is valid for {\em all} values of the matter mass terms $m$, including large values where SQCD becomes strongly coupled.
Our second important  results is the statement that the transition from locally to topologically distinguishable walls occurs right at the endpoint -- at infinite quark masses $m = \infty$. 
This is a subtle point because once the matter fields become heavy (but not infinitely heavy)
the low-energy world volume theory does not tell us that the degenerate walls are locally distinguishable. In fact, to see the difference between them one has to excite heavy modes inside the wall.

\section{SU(2) with one flavor}
\label{su21fla}

SU(2) is a quasi-real group and therefore one flavor can be split into two subflavor fields  $Q^f\,\, (f=1,2)$,  both in  the fundamental representation of 
SU(2)$_{\rm gauge}$, with the following Lagrangian for the matter sector
\begin{equation}
{\cal L}_{2,1} = \int d^{2}\theta d^{2}\bar{\theta}\,\bar{Q}^{\,f}e^{V}Q_{f}\,+
\left( \frac{m_0}{2}\int d^{2}\theta\,
	Q_{\alpha}^{f}Q_{\,f}^{\alpha}+{\rm H.c.}\right)
\end{equation}
Here $\alpha$ and $f$  are the color and flavor indices, respectively.  In what follows, it will be convenient to introduce the ``meson'' gauge invariant chiral superfield
\begin{equation}
	M = \frac{1}{2} Q_{\alpha}^{f}Q_{\,f}^{\alpha} \,.
\label{fourp}
\end{equation}
The first (kinetic) 
term is of the $D$ type, the second of the $F$ type; $m_0$ is the bare (UV) mass 
in the classical superpotential. The latter is not perturbatively renormalized. Note that $m_0$ cannot be set to zero because 
the vacua are run-away in this limit.

The  kinetic terms acquires a $Z\neq 1$ factor both from perturbative and nonperturbative contributions. These contributions are small at small $m_0$ since this is a weak coupling regime; they are expected to be small at large 
$m_0$ as well due to $\Lambda/m_0$ suppression ($\Lambda$ is the dynamical scale of the theory). We assume that $Z\sim 1$ and  nonsingular at
$m_0\sim\Lambda$.

\subsection{Superpotential and vacua}

The nonperturbative correction to the superpotential $F$ term 
was found by Affleck, Dine and Seiberg (ADS) leading to the famous ADS superpotential \cite{Affleck:1983mk},
%
\begin{equation}
{\cal L}_{\rm sp}=  \left( {m_0}\int d^{2}\theta\,
	M +\int\, d^2 \theta\, {\mathcal{W}}_{\rm{ADS}} \right), \quad
    \boxed{{\mathcal{W}}_{\rm{ADS}}}  = \frac{\Lambda^5}{M}, 
\label{2.3}
\end{equation}
see above for the definition of the dynamical scale $\Lambda$.
Of paramount importance are two facts: (i) using the generalized $R$ symmetry one can prove 
that the nonperturbative contribution in \eqref{2.3} is generated by one and only one instanton and is $m$ independent; 
(ii) using the supersymmetric instanton calculus \cite{Novikov:1985ic,Shifman:1999mv} one can prove that the instanton measure $d\mu$ is saturated at $\rho =0$ where $\rho$ is the instanton scale modulus. (The latter circumstance was not known to ADS.)
More exactly, 
\begin{multline}
	d\mu = \frac{1}{2^{5}} \frac{\Lambda^{5}}{M(x_0,\theta_{0})}
		\exp\left(-4\pi^{2} |M| \rho_{\mathrm{inv}}^{2}\right) \\
        \times
		\frac{\mathrm{d}\rho^{2}}{\rho^{2}}\,d^{4}x_{0}d^{2}\theta_{0}\,d^{2}\bar{\beta}\,d^{2}\bar{\theta}_{0}
\label{2.4}
\end{multline}
where the moduli $x_0,\,\theta_0\, ,\bar{\beta},\, \bar{\theta}_0, \, \rho$ are defined in \cite{Novikov:1985ic,Shifman:1999mv} and 
$\rho^2_{\rm inv}
=\rho^2  \left(1-4i \bar{\beta}\bar{\theta}_0\right)^{-1}$ 
is the superinvariant extension of the scale modulus $\rho^2$ obtained in the same work. Naively \eqref{2.4} vanishes after 
integrating over
$\bar{\beta}$, and $\bar{\theta}_{0}$. Indeed, these variables 
enter the integrand of the measure only through $\rho_{\mathrm{inv}}^{2}$. Replacing $d\rho^2/\rho^2$ by  $d\rho^2_{\rm inv}/\rho^2_{\rm inv}$
we are left with the vanishing Berezin integral over
$\bar{\beta}$, and $\bar{\theta}_{0}$.

The loophole is due to the singularity in the integrand at
$\rho_{\mathrm{inv}}^{2}=0$. To resolve the singularity we
integrate first over the fermionic variables and arrive at the identity
\begin{equation}
	\int\frac {d\rho^2}{\rho^2}\,d^{2}\bar{\beta}\,d^{2}\bar{\theta}_{0} F\left(\rho_{\rm inv}^{2}\right)
	= \int16 \, {d\rho_{\rm inv}^2}{\delta(\rho_{\rm inv}^{2}})F\left(\rho_{\rm inv}^{2}\right)
\label{2.5}
\end{equation}
which yields the superpotential $ {\mathcal{W}}_{\rm{Affleck:1983mk}} $ in \eqref{2.3}. 
Thus, the instanton calculation is {\em saturated} by the {\em zero-size} instanton. 
This observation was made in \cite{Novikov:1985ic,Shifman:1999mv} (see Sec. 4 there) long before the invention of the Nekrasov localization \cite{Nekrasov:2002qd}. 

The zero-size saturation is not accidental. It is related to the anomaly in the brane charge $\left[Z_{\alpha\beta}\right]_{\rm anom} \sim \lambda\lambda$
(see \eqref{two} and \cite{Dvali:1996xe,Shifman:2012zz}, Sect. 10.16.7). The zero-size saturation guarantees that \eqref{2.3} is valid for all values 
of the mass parameter $m_0$ -- small and large.

\subsection{Two locally distinguishable  domain walls}
\label{secthree}

\begin{figure}[ht]
    \subfloat[Small $m$]{\label{fig:quark_wall_weak_coupling}\includegraphics[width=0.3\columnwidth]{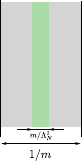}}
    \hspace{1cm}
    \subfloat[Large $m$]{\label{fig:quark_wall_strong_coupling}\includegraphics[width=0.3\columnwidth]{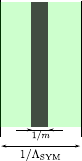}}
\caption{
	Domain wall structure.
	At small $m$ gluons are heavy, and the wall thickness is determined by the quark mass $\ell\sim 1/m$; inside there is a core of heavy gluons with mass determined by the matter fields VEVs, see  \eqref{3.3}.
	At large $m$ there is a thin quark core and a gluon cloud around it.
	}
\label{quark_wall_struct}
\end{figure}

The above consideration leads us to the overall picture depicted in Fig. \ref{quark_wall_struct}.
The domain walls in SQCD have two components -- one is built mainly of matter fields, the other from gluons (gluinos). At small $m\ll\Lambda$ all gauge bosons are Higgsed and heavy, they constitute a thin core with broad tails due to light matter fields, see Fig. \ref{quark_wall_struct}a.  In the opposite limit
$m\gg\Lambda$ the thin core is built of matter fields while the gluonic tails extend much further Fig. \ref{quark_wall_struct}b.
In both cases there are two wall trajectories which locally distinguish the wall from each other, see Fig. \ref{2w-junc}. In the former case the difference between the walls can be detected in the light (almost massless) modes, i.e. in the tails. In the latter case one has to excite the heavy modes to detect the distinctions in the wall core.


Field profiles on the wall can be found at small quark masses $m_0$.
For a domain wall interpolating between the vacua $\langle M\rangle= \pm (m_0)^{-1}\sqrt{m_0\Lambda^5}$  following from the superpotential \eqref{2.3}
the BPS wall equations are 
\begin{equation}
	\frac{\partial M}{\partial z}
		= \pm \frac{1}{2} |M| \frac{\partial \Big(\bar{\cal{W}}_{\rm cl}+ \bar{\cal{W}}_{\rm inst} \Big) }{\partial\bar {M}} \,.
\label{1five}
\end{equation}
The BPS wall solutions take the form (see Sec. 5 in \cite{Kovner:1997ca})
\begin{equation}
\begin{aligned}
	M_{\rm wall} &= \abs{\expval{M}} \, e^{\pm i\alpha(z)}\,,\\
	i\alpha(z) &= 2\log \left(\frac{1+i\,e^{m_0(z-z_0)}}{\sqrt{1+e^{2m_0(z-z_0)}}}\right)
\end{aligned}
\label{1six_p}
\end{equation}
with $|M_{\rm wall}| = |\expval{M}| ={\rm const}$ on the wall trajectory.
See Fig.~\ref{2w-junc} for a visual representation.

The solution \eqref{1six_p} is obtained in the quasiclassical limit of small $m_0$.
Unlike the BPS protected quantities the trajectory is subject to corrections; however, these corrections do not qualitatively change the wall profile even if $m_0$ becomes large. 
In equation \eqref{1six_p} the wall thickness if $\sim m_0^{-1}$. We do not see the core of the wall built of heavy fields --- Higgsed gauge bosons -- at small $m_0$.

Needless to say, to observe difference in the internal wall structure one must have {em probe} fields coupled to matter $M$ with an arbitrary weak interaction.
For large $m_0$ the core is made from matter and to see the difference between the two walls above we need to excite heavy modes.

\section{Arbitrary numbers of colors \boldmath{$N$} and flavors \boldmath{$F < N$}}
\label{sec3}

Let us continue this discussion directly in a more general case of $SU(N)$ SQCD with arbitrary number of colors $N$ and a number $F < N$ of matter multiplets $Q_f$ and $\tilde{Q}^{\bar{g}}$ in the (anti-)fundamental representation; these are quark flavors, with flavor indices $f, \bar{g} = 1, \ldots, F$.
For massless quarks the theory has runaway vacua, which are stabilized if we introduce the quark mass $m$  assuming  {\em all}  flavors to have the same $m$.\footnote{Strictly speaking, the weak coupling regime requires hierarchical masses. However, one can always rescale the matter fields 
to maintain flavor symmetry in the superpotential. This ruins the flavor symmetry in the kinetic term. The precise form of the latter is not important for the 
qualitative features and certainly does not change the CIVF index  \cite{Cecotti:1992qh} counting BPS walls.}
In the same way as in the $SU(2)$ SQCD case, the quark superpotential gets a contribution from instantons \cite{Affleck:1983mk}.
It is convenient to pass to the meson (gauge invariant) matrix
\begin{equation}
	M_f^{\bar{g}} = Q_f \tilde{Q}^{\bar{g}} \,, \quad
	f, \bar{g} = 1, \ldots, F \,.
\label{meson_field_def}
\end{equation}
The full superpotential for the theory with $N$ colors and $F$ flavors reads
\begin{equation}
	\mathcal{W} = m_0 \Tr M + (N - F) \frac{ \Lambda^{3N - F} }{ \det M } \,.
\label{quark_superpotential}
\end{equation}
As in Sec. \ref{su21fla}, this result is exact and valid for all values of $m$.
Here, $\Lambda$ is the strong coupling scale for the theory with $F$ quark flavors.
The theory has $N$ gapped vacua with VEVs
\begin{multline}
	\expval{M}_n = \frac{1}{m}\expval{  \frac{ \Tr \lambda\lambda }{16\pi^2} }_n  = \mathtt{M}_n \cdot \mathbb{I}_{F \times F} \,, \\
	\mathtt{M}_n =  e^{\frac{2 \pi i n}{N}} \cdot \left( \frac{ \Lambda^{3N - F} }{ m^{N - F} } \right)^{1/N}  \,, \quad n=1,..., N,
\label{3.3}
\end{multline}
where $\mathbb{I}_{F \times F}$ is an identity matrix.
The last formula follows from the Konishi relation \cite{Konishi:1983hf,Shifman:2012zz}.
The bare quark mass in these vacua is still $m_0$.
Leaving aside translational modulus and its superpartners we will focus on the reduced moduli space for $k$ walls $\widetilde{\mathcal{M}}_k$
and the corresponding world volume theory. It is natural to expect that this space is determined purely by the flavor symmetries broken by the walls.

A special case of particular interest $F=N$ was thoroughly analyzed\,\footnote{If $F=N$ the superpotential vanishes and a constraint on $M_f^{\bar{g}}$ arises as was first discussed by Seiberg \cite{Seiberg:1994bz}. On the nonbaryonic
Higgs branch it  takes the form det$M=\Lambda^{2N}$.}
in  \cite{Ritz:2002fm,Ritz:2004mp}. In this case all gauge bosons are Higgsed and the largest flavor moduli space emerges.
The world volume theory on the wall is an $\mathcal{N}=2$ sigma model with the Grassmannian  target space \cite{Ritz:2002fm}
\begin{equation}
	\widetilde{\mathcal{M}}_k = G(k,N) = \frac{ U(N) }{ U(k) \times U(N-k) } \,.
\label{3.4}
\end{equation}
Note that supersymmetry in the world volume theory is {\em enhanced} -- the number of the conserved supercharges is four \cite{Ritz:2004mp}.

The Witten index for sigma model \eqref{3.4}, which is given by the Euler characteristic of the target space, coincides with the multiplicity formula \eqref{three}.

In the general case $F<N$ we will reproduce the multiplicity \eqref{three} and formulate the wall world volume theory \eqref{domain_wall_theory} below by analyzing the 
bulk theory compactified on a cylinder $\mathbb{R}^3 \times \mathbb{S}^1$ (sometimes referred to as ``circle compactification'')
and exploiting  the counting method of \cite{Ritz:2002fm}. If $F<N-1$ some of the non-Abelian gauge bosons are not Higgsed; these are strongly coupled. 
Our usual strategy is to start from the weak coupling regime and then proceed to strong coupling. Putting the theory on a small-raduius cylinder we achieve 
this goal. 
We note the final formula for the domain wall multiplicity, \eqref{multiplicity_cylinder_full}, is valid for $F=N-1$ and $F=N$ as well.

\section{SYM on a cylinder: Moduli and symmetries}
\label{su21fl}

In this section we will derive a full generalization of the Acharya-Vafa formula \cite{Acharya:2001dz} while staying in field theory framework.
To this end we formulate the bulk SYM theory with the gauge group $SU(N)$ on a cylinder, i.e. compactify  $\mathbb{R}^4\to  \mathbb{R}^3 \times \mathbb{S}^1$\cite{Davies:1999uw,Davies:2000nw}.  
We start from the semiclassical regime in which the $\mathbb{S}^1$ circumference  $L\ll1/\Lambda$. 
Then all the gauge bosons {\em not} belonging to the
Cartan subalgebra  are Higgsed, while those in the Cartan subalgebra give rise to moduli. 
Namely, 
on the Coulomb branch, the 3D infrared effective field is organized in an  $(N\!-\!1)$-component dimensionless chiral superfield in a Cartan subalgebra of $\mathfrak{su}(N)$; the latter has dimension $N-1$,
\begin{equation}
    \vec{X}=\vec{x}+\sqrt{2}\boldsymbol{\theta}\vec{\lambda}+\boldsymbol{\theta}^2\vec{F} \,,\qquad   \vec{X} =\{X^1, X^2,... X^{N-1}\}\,.
\end{equation}
The scalar field $\vec{x}$ is the 3D reduction of the gauge field, i.e.,
\begin{equation}
    \vec{x}=\frac{8\pi^2}{Ng^2 }\,\vec{\rho} - \left(\frac{4\pi}{g^2}\, \vec{\sigma}  - \i\vec{\gamma}\right),\qquad \vec{\rho}\equiv\sum_{j=1}^{N-1}\vec{\mu}_j
\label{x_moduli_def}
\end{equation}
where $\vec{\mu}_j$'s are the simple weights and $\vec{\rho}$ is usually called the Weyl vector. Moreover,  $g$ is the coupling of the 3d theory.
The real part $\vec{\sigma}$ is the component of the gauge field along the compact direction $\mathbb{S}^1$, and the imaginary part $\vec{\gamma}$ is the dual gauge field, i.e. the dual scalar of the spatial gauge field.
They take values in an orbifold,
\begin{equation}
    (\vec{\sigma},\vec{\gamma})\in\left(\frac{\R^{N-1}}{2\pi\Gamma_r}\times\frac{\R^{N-1}}{2\pi\Gamma_w}  \right)\Bigg/W_{\mathfrak{su}(N)}
\label{monopole_moduli_space_p}
\end{equation}
where $\Gamma_r$, $\Gamma_w$, and $W_{\mathfrak{su}(N)}$ denote the root lattice, the weight lattice, and the Weyl group, respectively.

In 3D reduction, the theory has 0-form $\Z_{2N}$ chiral symmetry, a 0-form $\Z_N$ center symmetry, and a 1-form $\Z_N$ center symmetry.
The two 0-form symmetries act on $\vec{X}$ as follows,
\begin{subequations}
\begin{gather}
   \! \!\text{center: }\quad\vec{X}\to\vec{X}+2\pi\vec{\mu}\,,\quad \vec{\mu}\in\Gamma_w\!\!\!\mod\Gamma_r \,,\\
    \text{chiral: }\,\, \quad \vec{X}\to\vec{X}+\i\frac{2\pi k}{N}\vec{\rho}\,,\quad k\in\Z\!\!\!\mod 2N \,.
\end{gather}
\label{center_chiral_Xpp}
\end{subequations}
Recall that $\Gamma_w/\Gamma_r=\Z_N$.
The Coulomb branch is conveniently parameterized by ``monopole moduli'' \cite{Unsal:2007jx} defined as
\begin{equation}
    Y_j\equiv\exp{\vec{\alpha}_j \cdot\vec{X}}\,,\qquad j\in\Z\mod N
\label{Y_def}
\end{equation}
where $\vec{\alpha}_{j=1,\cdots,N-1}$'s are simple roots, and $$\vec{\alpha}_0=-\sum_{j=1}^{N-1}\vec{\alpha}_j$$ is the affine root.
Thus these $N$ operators are subject to a constraint
\begin{equation}
    \prod_{j\in\Z_N} Y_j = 1\,.
\label{Y_constraint_p}
\end{equation}
The monopoles corresponding to $Y_j$ with $j=1,\cdots,N-1$ are the usual (fundamental) monopoles, while the one corresponding to $Y_0$ with the affine root is the Kaluza-Klein (KK) monopole.

The actions of symmetry generators \eqref{center_chiral_Xpp} now take a very simple form:
\begin{subequations}
\begin{gather}
    \!\!\!\text{center: }\quad Y_j\to Y_{j+1}\,,\\[1mm]
    \text{chiral: }\quad Y_j\to Y_j\e^{\i\frac{2\pi}{N}}\,.
\end{gather}
\label{center_chiral_Yp}
\end{subequations}
The second line here follows directly from the second shift in  \eqref{center_chiral_Xpp}.
To understand the first line in  \eqref{center_chiral_Yp}, recall that the first shift in  \eqref{center_chiral_Xpp} gets us outside of the fundamental domain \eqref{monopole_moduli_space_p}, and to get back we need to perform a Weyl reflection; the resulting transformation amounts to a rotation.

The 1-form symmetry is realized as a solitonic symmetry acting on line defects with nontrivial winding number for $\vec{\gamma}$.
It is enhanced from $\Z_N$ to $U(1)^{N-1}$ on the Coulomb branch.

As pointed out by the authors of~\cite{Davies:1999uw,Davies:2000nw}, monopole-instantons generate the superpotential in the 3d EFT
\begin{equation}
	\mathcal{W}_{3d} = L\Lambda^3 \sum_{j\in\Z_N} Y_j \,.
\label{W_SYM_1p}
\end{equation}
Note that for SU(2) SYM, given \eqref{Y_constraint_p}, the superpotential \eqref{W_SYM_1p} functionally coincides with ${\cal L}_{\rm sp}$
in \eqref{2.3}, {\em with all ensuing consequences}. 

Equation  \eqref{W_SYM_1p} clearly respects the 0-form center symmetry and has a definite charge under the chiral symmetry.
Solving $\partial_{\vec{X}}\calW_{3d}=0$, we can find $N$ vacua parameterized by $n=1,\cdots,N$,
\begin{equation}
    \expval{\vec{X}}_n = -\i\frac{2\pi n}{N}\vec{\rho}\quad (\mathrm{mod}\ \i2\pi\Gamma_w)\,,\qquad \expval{Y_j}_n = \e^{-\i\frac{2\pi n}{N}}\,.
\end{equation}
The values of the superpotential in these vacua are given by
\begin{equation}
    (\calW_{3d})_n = L \Lambda^3 \e^{-\i\frac{2\pi n}{N}}\,.
\end{equation}
Clearly, chiral symmetry \eqref{center_chiral_Xpp} acts non-trivially on these vacua and center symmetry leaves them invariant, which signifies the chiral symmetry breaking $\Z_{2N}\to\Z_2$ and the confinement of the Polyakov loops (unbroken 0-form center symmetry).
The Wilson loops are also confined (unbroken 1-form center symmetry).

\section{Adding matter: from SYM to SQCD}
\label{six6}

When we introduce matter multiplets in the (anti-)fundamental representations, their scalar components can develop VEVs.
Generally speaking, this gives mass to the gauge fields and leads to confinement of the monopoles.

Which monopoles are affected upon introduction of fundamental matter? To answer this question it is helpful
to consider a generic scenario.
Let us start from a theory with a gauge group $G$.
After the circle compactification we are left with a 3d EFT where an Abelian gauge group $U(1)^r$ is unbroken by the adjoint VEVs (the Cartan torus of $G$, $r = \text{rank}(G)$).
Monopole charges $\vec{g}$ lie on the co-root lattice $\Gamma_r^*$; fundamental monopoles correspond to the simple roots, the KK monopole corresponds to the affine root.
Quarks that were in the fundamental representations of $G$ now acquire electric charges $\vec{Q}^\text{el}_j = \vec{w}_j$ that are in the weight lattice,  $\vec{w}_j \in \Gamma_w$.

Now suppose that a single quark $q$ with charge $\vec{Q}^\text{el}_j = \vec{w}_j$ acquires a scalar VEV $\expval{q} = v_j$.
All the monopoles with charges that have non-zero scalar product with $\vec{w}_j$ become confined, and there are exactly two such monopoles --- the one with magnetic charge $\vec{Q}^\text{magn}_j = \vec{\alpha}_j$ such that $\vec{w}_j \cdot \vec{\alpha}_j = 1$, and the KK monopole with charge $\vec{Q}^\text{magn}_0 = \vec{\alpha}_0 = - \sum \vec{\alpha}_j$ such that $\vec{w}_j \cdot \vec{\alpha}_0 = -1$.
To see that these monopoles are indeed confined, note that the potential from a free monopole of charge $\vec{Q}^\text{magn}$ is $A \sim \vec{Q}^\text{magn} / |x|$.
Because of the term $  |A q|^2$ in the Lagrangian ($q$ is the squark with a VEV $\expval{q} \neq 0$) the monopole action $S_\text{mon}$ diverges. 
The monopole weight in the path integral vanishes,
\begin{equation}
	e^{ - S_\text{mon} } = 0  \,.
\end{equation}
In a more general case when there are $F$ different quark flavors that develop VEVs, $F+1$ monopoles become confined.
We will see that the VEV $\expval{q}$ is indeed non-zero for quarks of finite mass, and vanishes only at the endpoint of infinite mass.
The corresponding U(1) gauge field mass is
$m_{{\rm U}(1)} = g_{3d} \expval{q_{3d}}$.

A pair of such monopoles with charges $\vec{\alpha}_j$ and $\vec{\alpha}_0$ form a confined object with a non-zero net magnetic charge 
\begin{equation}
	\vec{Q}_{0,j}^\text{magn} = \vec{\alpha}_j + \vec{\alpha}_0 = - \sum_{i \neq j}  \vec{\alpha}_i \,.
\end{equation}
Note that this $\vec{Q}_{0,j}^\text{magn}$ is orthogonal to the quark electric charge $\vec{Q}^\text{el}_j = \vec{w}_j$, which renders the action of this composite monopole finite.

In a more general scenario when we introduce $F$ flavors of quarks, the gauge symmetry is broken as
\begin{equation}
	SU(N) \xrightarrow{\text{ adj } \sigma \text{ } } U(1)^{N-1} \xrightarrow{\text{ fund } q \text{ } } U(1)^{N - 1 - F} \,.
\end{equation}
Correspondingly, $F+1$ out of $N$ monopoles become confined.
All these $F+1$ monopoles can form a free ``molecule'' with a nonvanishing net magnetic charge, generalizing the confined pair above.
This picture remains valid for $F < N-1$.
When the number of flavors reaches $F = N - 1$ or $F = N$, the gauge group is completely broken, and all the monopoles become confined.
The net magnetic charge of the ``molecule'' is now zero\footnote{The difference between $F = N - 1$ and $F = N$ in this context manifests itself in SQCD with gauge group $U(N)$ rather than $SU(N)$. In the former theory, at $F=N-1$ the monopole ``molecule'' resembles a chain with a non-zero leftover magnetic charge that is still unconfined, while at $F=N$ that leftover magnetic flux is also confined, and the ``monopole chain'' closes to a ``monopole necklace''.}.

Full superpotential can be derived from the path integral; the computation carried out in \cite{Csaki:2017mik,Shirman:2019mqv} accounts for the fact that with nonvanishing  quark VEVs, (some of the) monopoles become confined. Let us define
\begin{equation}
	Y_{\text{conf}} \equiv \frac{ [ \prod_{i=1}^{F+1} Y_i ] }{\det M} \,.
\label{Y_conf_def}
\end{equation}
where in the numerator we have a product of all the confined monopole moduli (recall that $Y$'s are given by an exponential of the adjoint VEVs, see \eqref{Y_def}).
Then the superpotential takes the form
\begin{multline}
    \mathcal{W} = \eta Y_0 + \sum_{i = F + 2}^{N - 1} Y_i + Y_{\text{conf}}  + m \Tr(M) \\
    - \lambda \left( \det(M) \cdot Y_0 \cdot Y_{\text{conf}} \cdot \prod_{i=F+2}^{N - 1} Y_i - 1 \right) \,.
\label{superpotential_3d_Nf_Yconf}
\end{multline}
Here, $\eta \equiv L^N \Lambda^{3N-F}$ is the 3d analog of the 4d one-instanton factor.
All  $N$ vacua can be readily found from \eqref{superpotential_3d_Nf_Yconf}.

\section{Domain walls at \boldmath{$0 \leqslant F \leqslant N$}}
\label{sec:walls}

In this Section we are going to apply our knowledge of the superpotentials of bulk SQCD compactified on a circle.
We are going to study the BPS domain wall trajectories, which will enable us to count the wall multiplicities and make a proposal for an effective theory living on the wall.

When the bulk theory is compactified on $\mathbb{R}^3 \times \mathbb{S}^1$ with circumference $L$, for self-consistency we require that the domain wall wraps the compact dimension.
Otherwise we would only be able to consider the wall-antiwall pairs, because the vacuum structure has to be periodic on $\mathbb{S}^1$.


Our starting point is the superpotential \eqref{superpotential_3d_Nf_Yconf}.

Take a $k$-wall interpolating between vacua $n=n_0$ and $n= n_0 + k$ (we will focus on $0< k \leqslant N/2$).
We denote eigenvalues of the meson matrix $M$ by $x_i$, $i=1,\ldots,F$.
The constraint encoded by the superpotential \eqref{superpotential_3d_Nf_Yconf} reads
\begin{equation}
	\left( \prod_{i=1}^F x_i \right) \cdot Y \cdot Y_{\text{conf}} \cdot \left( \prod_{i=F+2}^{N - 1} Y_i \right) = 1 \,.
\label{F_constraint}
\end{equation}
Thus, for any $F$, we have a product of $N$ complex variables, constrained to be 1. If $F=N$ then we would have $\prod_i^N x_i =1$. The monopole moduli 
$Y$ provide the missing $N-F$ factors in \eqref{F_constraint}. This fact allows us to proceed in the same vein as in the $N=F$ case, see \cite{Ritz:2002fm}.

Thus, the generic domain wall counting is essentially a crossbreed between the topological Chern-Simons formula of Acharya and Vafa
and non-linear sigma model (NLSM) analysis  of \cite{Ritz:2002fm}.

\subsection{Analysis of the BPS wall trajectories}

\begin{figure}[h]
    \centering
    \includegraphics[width=0.3\textwidth]{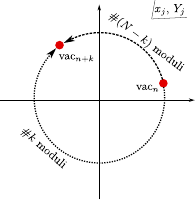}
    \caption{
    	Winding of the bulk moduli inside a BPS domain wall.
    	The bulk moduli are the eigenvalues of the meson matrix $x_j$ ($F$ of them, where $F$ is the number of quark flavors) and/or the monopole moduli $Y_j$ ($N-F$ of them for the gauge group $SU(N)$).
    	The red dots represent two chiral vacua.
    	}
    \label{fig:moduli_windings}
\end{figure}

We want to study more closely the problem of the domain wall counting based on the superpotential \eqref{superpotential_3d_Nf_Yconf} and the corresponding constraint \eqref{F_constraint}.
In total we have $N$ complex variables: $F$ of them are flavor moduli $x_i$ and the rest $N-F$ are the monopole moduli $Y_0, Y_\text{conf}, Y_i$.
Combining the both lines of rasoning we conclude that for a $k$-wall
exactly $k$ out of these $N$ variables have to wind clockwise, while the rest $N-k$ wind counter-clockwise, see Fig.~\ref{fig:moduli_windings}.

To distinguish between local and topological contributions to the multiplicity, we actually need to keep a closer look at windings of the monopole and flavor moduli separately.
{\it Let $J$ be the number of flavor moduli that wind clockwise}.

Obviously $J$ cannot be larger than the total number of flavors $F$, and because exactly $k$ out of $N$ complex variables wind clockwise, $J$ cannot exceed $k$; thus $J \leqslant \min(k,F)$.
On the other hand, $J$ must be non-negative, and because exactly $N-k$ variables wind counter-clockwise ($F-J$ of those are the flavor moduli $x_i$) we have $N-k \geqslant F-J$; thus $J \geqslant \max(0,F+k-N)$.
All in all, the allowed values of $J$ lie in the interval
\begin{equation}
	\max(0,F + k - N) \leqslant J \leqslant \min(k,F) \,.
\label{J_interval}
\end{equation}

Now we are ready to write down the formula for domain wall multiplicity.
In a sector with a given $J$, we have $C^F_J$ choices to pick which of the flavor moduli wind clockwise, and on top of that we also have $C^{N-F}_{k-J}$ choices to pick which of the monopole moduli wind clockwise.
Once these choices are made, the rest of the moduli are bound to wind in the opposite direction.
Summing up over the range of possible values of $J$ gives the total multiplicity of $k$-walls
\begin{equation}
    \nu_{N,k}^\text{walls} \Big|_{\mathbb{R}^3 \times \mathbb{S}^1} = \sum_{J=\max(0,F + k - N)}^{ \min(k,F) } 
    C^{N-F}_{k-J} C^F_J = C^N_k
\label{multiplicity_cylinder_full}
\end{equation}
which of course reproduces\footnote{As was noted in footnote 13 of \cite{Bashmakov:2018ghn}, the last equality sign in \eqref{multiplicity_cylinder_full} can be proven by applying the binomial expansion formula on the two sides of the identity $(1+t)^{N-F} (1+t)^F = (1+t)^{N}$.} 
\eqref{three} protected by the Witten index.
Conceptually, the multiplicity formula \eqref{multiplicity_cylinder_full} is similar to the CFIV index \cite{Cecotti:1992qh}.
Multiplicity formula \eqref{multiplicity_cylinder_full} was derived previously in \cite{Bashmakov:2018ghn} from an entirely different argument.

We can argue that, in the decompactified bulk SQCD on $\mathbb{R}^4$, the domain wall effective theory is a product\footnote{In \eqref{domain_wall_theory}, in the two-level notation $U(a)_{\ell_1,\ell_2}$ the first $\ell_1$ is the level of the non-Abelian factor $SU(a)$ while $\ell_2$ is the level of the Abelian factor $U(1)$.}
\begin{equation}
\boxed{
\begin{aligned}
DW_{\rm worldvolume}= {\rm U}&(k-J)_{N-F-k+J,\, N-F} \, \text{ CS} \\  
&\times  \frac{ {\rm U}(F) }{ {\rm U}(J) \times {\rm U}(F-J) } \text{ NLSM} \,.  
\end{aligned}
}
\label{domain_wall_theory}
\end{equation}
This is our hypothesis for the effective theory on the domain wall for \textit{any} finite value of the quark mass.
Note that in the case of pure SYM with $F=0$ \eqref{domain_wall_theory} reproduces the pure Chern-Simons theory of Acharya and Vafa \cite{Acharya:2001dz}.
When one introduces quark flavors, the corresponding flavor moduli form a sigma model on the wall and at the same time confine some of the monopole moduli thus ``eating up'' part of the CS theory resulting in \eqref{domain_wall_theory}.
We note also that \eqref{domain_wall_theory} respects the parity turning a $k$-wall into an $(N-k)$-wall (this symmetry redices to the level-rank duality in the CS sector of the theory).

We stress that the multiplicity formula \eqref{multiplicity_cylinder_full} obtained from combinatorics of the wall trajectories is valid for any number of quark flavors in the interval $0 \leqslant F \leqslant N$. 
Although the argument above does not cover $F=N-1$ and $F=N$ strictly speaking, it can be generalized to include these cases as well \cite{Chen:2024jyn}, see also here below.
In particular, for $F=N$ there are no monopole moduli so that $J \equiv k$ for all walls, while in the pure SYM case there are no flavor moduli and $J \equiv 0$.

\subsection{Local and topological multiplicity}
\label{sec:local_top_mult_discuss}

To back up our formula \eqref{domain_wall_theory} let us derive \eqref{multiplicity_cylinder_full} starting from the world volume formula \eqref{domain_wall_theory}.

First of all, the Witten index for the $\mathbb{G}\mathrm{r}(J,F)$ sigma model in \eqref{domain_wall_theory} is
\begin{equation}
I_W= C^F_J
\end{equation}
which corresponds to the second factor in the sum in \eqref{multiplicity_cylinder_full}.
Witten index in this sector remains well-defined after decompactification because the matter sector is IR regularized by the mass term.
This part of the multiplicity is due to the flavor degrees of freedom. 
Different domain walls in this sector can be distinguished locally, and junctions between them are possible (see more on that below).
On the wall effective theory, these junctions correspond to solitons of the sigma model interpolating between different ground states.
This motivates us to call this contribution \textit{local} multiplicity.

On the contrary, the first factor in the sum in  \eqref{multiplicity_cylinder_full} is due to a very different physics of the gauge fields.
The Witten index for the $U(k-J)_{(N-k) - (F-J), N-F}$ CS TQFT from  \eqref{domain_wall_theory} is well-defined only when the theory is compactified on a torus or a cylinder \cite{Acharya:2001dz} (the size of the compact dimension(s) may be large but finite).
Only in that case we see the corresponding contribution $C^{N-F}_{k-J} $ to the wall multiplicity%

On the wall world volume, this index counts the number of ground states of this TQFT; however, two wall junctions between them do not exist.
Moreover, if we decompactify completely and return to the $\mathbb{R}^4$ bulk, this multiplicity disappears; the Witten index is not well-defined in this case.
This motivates us to call this contribution \textit{topological} multiplicity.

Note that formulas \eqref{domain_wall_theory} and \eqref{multiplicity_cylinder_full} work even for the borderline cases of $F=N-1$ and $F=N$.
The latter is  obvious; the former is illustrated by the example $N=2$ and $F=1$; this corresponds to the SQCD discussed in Secs.~\ref{2w-junc}, \ref{secthree} and we expect to find two domain walls with no non-trivial moduli\footnote{Except the translational moduli which we do not discuss here, since they are free and decoupled.} and with a possible two-wall junction.
Indeed, the multiplicity formula \eqref{multiplicity_cylinder_full} gives
\begin{equation}
	  \begin{pmatrix} 1 \\ 1 \end{pmatrix}  \begin{pmatrix} 1 \\ 0 \end{pmatrix} 
	+ \begin{pmatrix} 1 \\ 0 \end{pmatrix}  \begin{pmatrix} 1 \\ 1 \end{pmatrix} 
	= 2 \,.
\end{equation}
The two walls on Fig.~\ref{2w-junc} correspond to $J=0$ for the upper one while $J=1$ for the lower wall.
The formula \eqref{domain_wall_theory} gives a trivial theory for each of these walls.

\subsection{Two-wall junctions: an example }
\label{sec:two_wall_junc}

We consider only the simplest example:  SU(2) and $F=2$ quark flavors on the $\mathbb{R}^4$ bulk.
In this theory there are two chiral vacua and two distinct domain walls interpolating between them, Secs.~\ref{2w-junc}, \ref{secthree}.
The two walls can form a junction, see Fig.~\ref{2w-junc}.
The effective theory on the wall world volume is 3d $\mathbb{CP}(1)$ model (this was also derived in \cite{Ritz:2002fm}), and the junction represents a domain line of this theory.

Let $m_1$, $m_2$ be the quark masses.
In the hierarchical limit $m_2 \gg m_1$ the effective $\mathbb{CP}(1)$ sigma model is weakly coupled, and the junction tension \cite{Ritz:2002fm} reads
\begin{equation}
	T_{junc} = \pi\left|\frac{\left|m_1\right|-\left|m_2\right|}{\sqrt{\left|m_1 m_2\right|}}\right| \Lambda_{F=2}^2
	\xrightarrow{m_2 \to \infty}
	\pi \sqrt{ \frac{\Lambda_{F=1}^5}{|m_1|} } \,.
\label{junc_tension_weak_1}
\end{equation}
When taking $m_2$ to infinity one should keep $\Lambda_{F=1}^5 = m_2 \Lambda_{F=2}^4$ fixed.

When we compactify the bulk to a cylinder $\mathbb{R}^3 \times \mathbb{S}^1$, the domain wall also wraps the compact dimension.
Consider the case when the junction also wraps the compact dimension.
In this case, the domain wall theory reduces to 2d $\mathbb{CP}(1)$ model, while the junction becomes a kink interpolating between two vacua (i.e. the two distinct domain walls).
The dynamical scale generated in the 2d effective theory is
\begin{equation}
	\Lambda_{2d} = \mu \, \exp( - \frac{2\pi}{g_{2d}^2(\mu)} )	\,, \quad
	\frac{1}{g_{2d}^2(\mu)} = L \, \frac{\Lambda_{4d}^2}{ \sqrt{m_1 m_2} } \,.
\label{2d_scale}
\end{equation}
This scale becomes relevant in the strong coupling regime $m_1 \sim m_2$, and we can make only some qualitative statements.
Here, $\mu$ is the UV scale of the effective theory set by the bulk quark masses.

At weak coupling the kink mass in the 2d $\mathbb{CP}(1)$ model is given by
\begin{equation}
	M_{kink} \approx \frac{ |\Delta m| }{ g_{2d}^2 }
\label{2d_kink_mass_weak}
\end{equation}
where $\Delta m = m_2 - m_1$ is the mass scale of the theory.
At small $\Delta m$ the theory becomes strongly coupled, and the kink mass is given by the central charge \cite{Dorey:1998yh}, which at $\Delta m = 0$ reads
\begin{equation}
	M_{kink} = \frac{1}{ 2 \pi e }  \Lambda_{2d} \,.
\label{2d_kink_mass_strong}
\end{equation}
where $e = 2.718 \ldots$ is the Euler's number.

To see the consistency between these results, note that when the junction world line wraps the compact dimension, the junction tension and the mass of the corresponding kink in the 2d effective theory on the domain wall are related as
\begin{equation}
	M_{kink} = L \, T_{junc} \,.
\label{kink_junc_mass_relation}
\end{equation}
Then, at weak coupling \eqref{junc_tension_weak_1} and \eqref{kink_junc_mass_relation} together give \eqref{2d_kink_mass_weak}.
At strong coupling, $g_{2d}^2$ becomes large, and \eqref{2d_scale} and \eqref{2d_kink_mass_strong} imply that the 2d kinks become heavy, with masses of the order of the bulk quark mass.
This is expected behavior agreeing with the comment at the end of Sec.~IV.B of \cite{Ritz:2004mp}.

\begin{figure}[h]
    \centering
    \includegraphics[width=0.5\columnwidth]{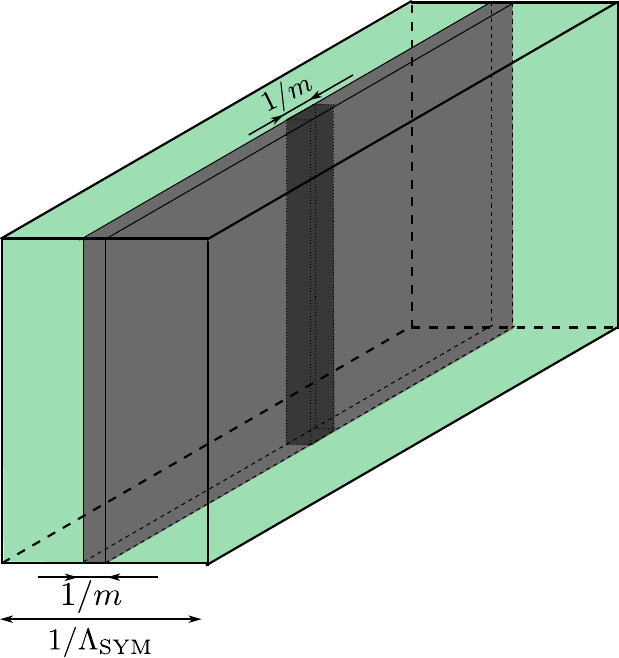}
    \caption{
    	Schematic structute of the two-wall junction for heavy quarks, $m \gg \Lambda$.
        The quark core (shaded) has different structure inside the two walls, while the gluon clouds (green) are nearly identical.
        The transition region of the junction is localized on the core (darkened region).
    }
    \label{fig:thick_junction}
\end{figure}%

We end this subsection with the following note.
The supersymmetric 2d $\mathbb{CP}(1)$ sigma model at $\Delta m = 0$ has two degenerate kink supermultiplets corresponding to the CFIV index = 2 \cite{Cecotti:1992qh}. 
The corresponding kinks have the same topological charge 1 and form a doublet with respect to the SU(2) flavor symmetry of this sigma model (see e.g. \cite{Witten:1978bc,Hori:2000kt} and references therein).
This suggests that the junction of the 4d SQCD in this case also form a representation with respect to the flavor symmetry.

Away from the point $\Delta m = 0$ flavor symmetry is broken, and the junction tensions are no longer degenerate.
In particular at large $\Delta m$ the theory flows to the SU(2) SQCD with $F=1$ flavor.
In this case the two kinks mentioned above have very different masses, only one of them being relatively light.
Precisely this light kink is seen as the junction in the 4d SQCD \cite{Ritz:2004mp}: in this case, the one flavor that we have interpolates between going clockwise in the first wall and going counter-clockwise in the second wall, see Fig.~\ref{2w-junc}.

\section{``Local-to-topological'' phase transition}

As was discussed above, domain walls in SQCD with light quarks are characterized by ``local'' multiplicity.
On the contrary, in pure SYM without quarks the domain wall multiplicity is ``topological''.
If we start with SQCD and slowly vary the quark mass parameter $m$, at some point we can expect to encounter some sort of a transition between these two regimes.
The question is, what sort of a transition is it, and where does it happen.

First of all, note that when the quark mass $m$ is small, the flavor symmetry is spontaneously broken on the wall.
The sigma model living on the wall worldvolume has multiple degenerate ground states.
On the other hand, in pure SYM the domain wall theory is a TQFT with a single ground state (when the bulk is taken to be the uncompactified $\mathbb{R}^4$.

This suggests the following scenario.
There is a phase transition associated with spontaneous breaking of the 0-form flavor symmetry on the wall world volume.
In SQCD with massive quarks this symmetry is spontaneously broken by the wall at any $m$.
However at the ``endpoint'' of infinite $m$, when we recover pure SYM, the flavor symmetry ceases to exist (there are no flavors left).
Therefore we argue that if we start in SQCD with locally distinguishable domain walls and start increasing the quark mass, we have a phase transition at the endpoint $m = \infty$ where the domain wall multiplicity becomes topological.

Thus in SQCD with quarks of any finite mass $m$, the domain walls are distinguishable by local experiments.
However, when the quarks are heavy, one might have to excite high-energetic modes of the wall to detect the quark core, see Fig. \ref{fig:thick_junction}.
Therefore, although the walls are ``local'', an observer constrained to low energies will perceive the walls as ``topological''.
In terms of the sigma model for the wall flavor moduli (see \eqref{domain_wall_theory}), the would-be quasi-massless modes become heavy.

\vspace{10pt}

To conclude, let us comment on one particular case.
Consider the $SU(N)$ SQCD on $\mathbb{R}^4$ (no compactification) and $F=N$ flavors of quarks with exactly identical masses $m$.
We'll focus on a minimal ($k=1$) domain wall in this theory.
At finite equal quark masses, such a wall supports a $\mathbb{CP}(N-1)$ model localized on the wall worldvolume $\mathbb{R}^3$, which is gapless. 
However, at $m=\infty$ we should recover SYM with a gapped domain wall theory (Yang-Mills + Chern-Simons). 
The question is, how a gapless theory becomes gapped?

The answer is somewhat peculiar.
When the bulk quark mass goes to infinity, $m \to \infty$, the $\mathbb{CP}(N-1)$ coupling on the wall vanishes.
The massless moduli of the quark wall simply become free fields supported on a quark core of thickness $1/m$ (dark grey slice on Fig.~\ref{fig:thick_junction}).
In the limit $m \to \infty$ this quark core becomes infinitely thin, and the wall effective theory is determined only by the gluon degrees of freedom that give the CS theory.

In the bulk this effect is seen as follows.
At finite quark masses the theory possesses $U(N)$ global flavor symmetry.
The massless moduli living on the wall are the Goldstones of this symmetry.
In the limit $m \to \infty$ the bulk quarks decouple, and, as a result, the $U(N)$ flavor transformations become trivial (not faithful).


\section{Acknowledgements} 
We thank Aleksey Cherman, Jaume Gomis, Zohar Komargodski, Sahand Seifnashri, Yue Zhao and Sergei Dubovsky for helpful discussions. This work is supported in part by U.S. Department of Energy Grant No. de-sc0011842 and the Simons Foundation award number 994302 (S.C.).

\bibliography{main}

\begin{thebibliography}{29}%
\makeatletter
\providecommand \@ifxundefined [1]{%
 \@ifx{#1\undefined}
}%
\providecommand \@ifnum [1]{%
 \ifnum #1\expandafter \@firstoftwo
 \else \expandafter \@secondoftwo
 \fi
}%
\providecommand \@ifx [1]{%
 \ifx #1\expandafter \@firstoftwo
 \else \expandafter \@secondoftwo
 \fi
}%
\providecommand \natexlab [1]{#1}%
\providecommand \enquote  [1]{``#1''}%
\providecommand \bibnamefont  [1]{#1}%
\providecommand \bibfnamefont [1]{#1}%
\providecommand \citenamefont [1]{#1}%
\providecommand \href@noop [0]{\@secondoftwo}%
\providecommand \href [0]{\begingroup \@sanitize@url \@href}%
\providecommand \@href[1]{\@@startlink{#1}\@@href}%
\providecommand \@@href[1]{\endgroup#1\@@endlink}%
\providecommand \@sanitize@url [0]{\catcode `\\12\catcode `\$12\catcode
  `\&12\catcode `\#12\catcode `\^12\catcode `\_12\catcode `\%12\relax}%
\providecommand \@@startlink[1]{}%
\providecommand \@@endlink[0]{}%
\providecommand \url  [0]{\begingroup\@sanitize@url \@url }%
\providecommand \@url [1]{\endgroup\@href {#1}{\urlprefix }}%
\providecommand \urlprefix  [0]{URL }%
\providecommand \Eprint [0]{\href }%
\providecommand \doibase [0]{https://doi.org/}%
\providecommand \selectlanguage [0]{\@gobble}%
\providecommand \bibinfo  [0]{\@secondoftwo}%
\providecommand \bibfield  [0]{\@secondoftwo}%
\providecommand \translation [1]{[#1]}%
\providecommand \BibitemOpen [0]{}%
\providecommand \bibitemStop [0]{}%
\providecommand \bibitemNoStop [0]{.\EOS\space}%
\providecommand \EOS [0]{\spacefactor3000\relax}%
\providecommand \BibitemShut  [1]{\csname bibitem#1\endcsname}%
\let\auto@bib@innerbib\@empty
\bibitem [{\citenamefont {Golfand}\ and\ \citenamefont
  {Likhtman}(1971)}]{Golfand:1971iw}%
  \BibitemOpen
  \bibfield  {author} {\bibinfo {author} {\bibfnamefont {Y.~A.}\ \bibnamefont
  {Golfand}}\ and\ \bibinfo {author} {\bibfnamefont {E.~P.}\ \bibnamefont
  {Likhtman}},\ }\bibfield  {title} {\bibinfo {title} {{Extension of the
  Algebra of Poincare Group Generators and Violation of p Invariance}},\ }\href
  {https://doi.org/10.1142/9789814542340_0001} {\bibfield  {journal} {\bibinfo
  {journal} {JETP Lett.}\ }\textbf {\bibinfo {volume} {13}},\ \bibinfo {pages}
  {323} (\bibinfo {year} {1971})},\ \bibinfo {note} {[Reprinted in G. Kane and
  M.~Shifman, {\sl The Supersymmetric World: The Beginnings of the Theory},
  Second Edition (World Scientific, Singapore, 2025), page 478]}\BibitemShut
  {NoStop}%
\bibitem [{\citenamefont {Wess}\ and\ \citenamefont
  {Zumino}(1974)}]{Wess:1974tw}%
  \BibitemOpen
  \bibfield  {author} {\bibinfo {author} {\bibfnamefont {J.}~\bibnamefont
  {Wess}}\ and\ \bibinfo {author} {\bibfnamefont {B.}~\bibnamefont {Zumino}},\
  }\bibfield  {title} {\bibinfo {title} {{Supergauge Transformations in
  Four-Dimensions}},\ }\href {https://doi.org/10.1016/0550-3213(74)90355-1}
  {\bibfield  {journal} {\bibinfo  {journal} {Nucl. Phys. B}\ }\textbf
  {\bibinfo {volume} {70}},\ \bibinfo {pages} {39} (\bibinfo {year}
  {1974})}\BibitemShut {NoStop}%
\bibitem [{\citenamefont {Bogomolny}(1976)}]{Bogomolny:1975de}%
  \BibitemOpen
  \bibfield  {author} {\bibinfo {author} {\bibfnamefont {E.~B.}\ \bibnamefont
  {Bogomolny}},\ }\bibfield  {title} {\bibinfo {title} {{Stability of Classical
  Solutions}},\ }\href@noop {} {\bibfield  {journal} {\bibinfo  {journal} {Sov.
  J. Nucl. Phys.}\ }\textbf {\bibinfo {volume} {24}},\ \bibinfo {pages} {449}
  (\bibinfo {year} {1976})},\ \bibinfo {note} {reprinted in C. Rebbi and G.
  Soliani (eds.), {\sl Solitons and Particles}, (World Scientific, Singapore,
  1984) p. 530}\BibitemShut {NoStop}%
\bibitem [{\citenamefont {Prasad}\ and\ \citenamefont
  {Sommerfield}(1975)}]{Prasad:1975kr}%
  \BibitemOpen
  \bibfield  {author} {\bibinfo {author} {\bibfnamefont {M.~K.}\ \bibnamefont
  {Prasad}}\ and\ \bibinfo {author} {\bibfnamefont {C.~M.}\ \bibnamefont
  {Sommerfield}},\ }\bibfield  {title} {\bibinfo {title} {{An Exact Classical
  Solution for the 't Hooft Monopole and the Julia-Zee Dyon}},\ }\href
  {https://doi.org/10.1103/PhysRevLett.35.760} {\bibfield  {journal} {\bibinfo
  {journal} {Phys. Rev. Lett.}\ }\textbf {\bibinfo {volume} {35}},\ \bibinfo
  {pages} {760} (\bibinfo {year} {1975})},\ \bibinfo {note} {reprinted in C.
  Rebbi and G. Soliani (eds.), {\sl Solitons and Particles}, (World Scientific,
  Singapore, 1984) p. 530}\BibitemShut {NoStop}%
\bibitem [{\citenamefont {Witten}(1982)}]{Witten:1982df}%
  \BibitemOpen
  \bibfield  {author} {\bibinfo {author} {\bibfnamefont {E.}~\bibnamefont
  {Witten}},\ }\bibfield  {title} {\bibinfo {title} {{Constraints on
  Supersymmetry Breaking}},\ }\href
  {https://doi.org/10.1016/0550-3213(82)90071-2} {\bibfield  {journal}
  {\bibinfo  {journal} {Nucl. Phys. B}\ }\textbf {\bibinfo {volume} {202}},\
  \bibinfo {pages} {253} (\bibinfo {year} {1982})},\ \bibinfo {note} {reprinted
  in S. Ferrara (ed.) {\sl Supersymmetry}, (North Holland/World Scientific)
  Amsterdam-Singapore, 1987, Vol. 1, page 490}\BibitemShut {NoStop}%
\bibitem [{\citenamefont {Dvali}\ and\ \citenamefont
  {Shifman}(1997)}]{Dvali:1996xe}%
  \BibitemOpen
  \bibfield  {author} {\bibinfo {author} {\bibfnamefont {G.~R.}\ \bibnamefont
  {Dvali}}\ and\ \bibinfo {author} {\bibfnamefont {M.~A.}\ \bibnamefont
  {Shifman}},\ }\bibfield  {title} {\bibinfo {title} {{Domain walls in strongly
  coupled theories}},\ }\href {https://doi.org/10.1016/S0370-2693(97)00131-7}
  {\bibfield  {journal} {\bibinfo  {journal} {Phys. Lett. B}\ }\textbf
  {\bibinfo {volume} {396}},\ \bibinfo {pages} {64} (\bibinfo {year} {1997})},\
  \bibinfo {note} {[Erratum: Phys.Lett.B 407, 452 (1997)]},\ \Eprint
  {https://arxiv.org/abs/hep-th/9612128} {arXiv:hep-th/9612128} \BibitemShut
  {NoStop}%
\bibitem [{\citenamefont {Kovner}\ \emph {et~al.}(1997)\citenamefont {Kovner},
  \citenamefont {Shifman},\ and\ \citenamefont {Smilga}}]{Kovner:1997ca}%
  \BibitemOpen
  \bibfield  {author} {\bibinfo {author} {\bibfnamefont {A.}~\bibnamefont
  {Kovner}}, \bibinfo {author} {\bibfnamefont {M.~A.}\ \bibnamefont
  {Shifman}},\ and\ \bibinfo {author} {\bibfnamefont {A.~V.}\ \bibnamefont
  {Smilga}},\ }\bibfield  {title} {\bibinfo {title} {{Domain walls in
  supersymmetric Yang-Mills theories}},\ }\href
  {https://doi.org/10.1103/PhysRevD.56.7978} {\bibfield  {journal} {\bibinfo
  {journal} {Phys. Rev. D}\ }\textbf {\bibinfo {volume} {56}},\ \bibinfo
  {pages} {7978} (\bibinfo {year} {1997})},\ \Eprint
  {https://arxiv.org/abs/hep-th/9706089} {arXiv:hep-th/9706089} \BibitemShut
  {NoStop}%
\bibitem [{\citenamefont {Witten}(1997)}]{Witten:1997ep}%
  \BibitemOpen
  \bibfield  {author} {\bibinfo {author} {\bibfnamefont {E.}~\bibnamefont
  {Witten}},\ }\bibfield  {title} {\bibinfo {title} {{Branes and the dynamics
  of QCD}},\ }\href {https://doi.org/10.1016/S0550-3213(97)00648-2} {\bibfield
  {journal} {\bibinfo  {journal} {Nucl. Phys. B}\ }\textbf {\bibinfo {volume}
  {507}},\ \bibinfo {pages} {658} (\bibinfo {year} {1997})},\ \Eprint
  {https://arxiv.org/abs/hep-th/9706109} {arXiv:hep-th/9706109} \BibitemShut
  {NoStop}%
\bibitem [{\citenamefont {Acharya}\ and\ \citenamefont
  {Vafa}(2001)}]{Acharya:2001dz}%
  \BibitemOpen
  \bibfield  {author} {\bibinfo {author} {\bibfnamefont {B.~S.}\ \bibnamefont
  {Acharya}}\ and\ \bibinfo {author} {\bibfnamefont {C.}~\bibnamefont {Vafa}},\
  }\href@noop {} {\bibinfo {title} {{On domain walls of N=1 supersymmetric
  Yang-Mills in four-dimensions}}} (\bibinfo {year} {2001}),\ \Eprint
  {https://arxiv.org/abs/hep-th/0103011} {arXiv:hep-th/0103011} \BibitemShut
  {NoStop}%
\bibitem [{\citenamefont {Ritz}\ \emph {et~al.}(2002)\citenamefont {Ritz},
  \citenamefont {Shifman},\ and\ \citenamefont {Vainshtein}}]{Ritz:2002fm}%
  \BibitemOpen
  \bibfield  {author} {\bibinfo {author} {\bibfnamefont {A.}~\bibnamefont
  {Ritz}}, \bibinfo {author} {\bibfnamefont {M.}~\bibnamefont {Shifman}},\ and\
  \bibinfo {author} {\bibfnamefont {A.}~\bibnamefont {Vainshtein}},\ }\bibfield
   {title} {\bibinfo {title} {{Counting domain walls in N=1 superYang-Mills}},\
  }\href {https://doi.org/10.1103/PhysRevD.66.065015} {\bibfield  {journal}
  {\bibinfo  {journal} {Phys. Rev. D}\ }\textbf {\bibinfo {volume} {66}},\
  \bibinfo {pages} {065015} (\bibinfo {year} {2002})},\ \Eprint
  {https://arxiv.org/abs/hep-th/0205083} {arXiv:hep-th/0205083} \BibitemShut
  {NoStop}%
\bibitem [{\citenamefont {Chen}\ \emph {et~al.}(2024)\citenamefont {Chen},
  \citenamefont {Ievlev},\ and\ \citenamefont {Shifman}}]{Chen:2024jyn}%
  \BibitemOpen
  \bibfield  {author} {\bibinfo {author} {\bibfnamefont {S.}~\bibnamefont
  {Chen}}, \bibinfo {author} {\bibfnamefont {E.}~\bibnamefont {Ievlev}},\ and\
  \bibinfo {author} {\bibfnamefont {M.}~\bibnamefont {Shifman}},\ }\href@noop
  {} {\bibinfo {title} {{Two types of domain walls in $\mathcal{N}=1$
  super-QCD: how they are classified and counted}}} (\bibinfo {year} {2024}),\
  \Eprint {https://arxiv.org/abs/2411.16845} {arXiv:2411.16845 [hep-th]}
  \BibitemShut {NoStop}%
\bibitem [{\citenamefont {Affleck}\ \emph {et~al.}(1984)\citenamefont
  {Affleck}, \citenamefont {Dine},\ and\ \citenamefont
  {Seiberg}}]{Affleck:1983mk}%
  \BibitemOpen
  \bibfield  {author} {\bibinfo {author} {\bibfnamefont {I.}~\bibnamefont
  {Affleck}}, \bibinfo {author} {\bibfnamefont {M.}~\bibnamefont {Dine}},\ and\
  \bibinfo {author} {\bibfnamefont {N.}~\bibnamefont {Seiberg}},\ }\bibfield
  {title} {\bibinfo {title} {{Dynamical Supersymmetry Breaking in
  Supersymmetric QCD}},\ }\href {https://doi.org/10.1016/0550-3213(84)90058-0}
  {\bibfield  {journal} {\bibinfo  {journal} {Nucl. Phys. B}\ }\textbf
  {\bibinfo {volume} {241}},\ \bibinfo {pages} {493} (\bibinfo {year}
  {1984})}\BibitemShut {NoStop}%
\bibitem [{\citenamefont {Novikov}\ \emph {et~al.}(1985)\citenamefont
  {Novikov}, \citenamefont {Shifman}, \citenamefont {Vainshtein},\ and\
  \citenamefont {Zakharov}}]{Novikov:1985ic}%
  \BibitemOpen
  \bibfield  {author} {\bibinfo {author} {\bibfnamefont {V.~A.}\ \bibnamefont
  {Novikov}}, \bibinfo {author} {\bibfnamefont {M.~A.}\ \bibnamefont
  {Shifman}}, \bibinfo {author} {\bibfnamefont {A.~I.}\ \bibnamefont
  {Vainshtein}},\ and\ \bibinfo {author} {\bibfnamefont {V.~I.}\ \bibnamefont
  {Zakharov}},\ }\bibfield  {title} {\bibinfo {title} {{Supersymmetric
  Instanton Calculus (Gauge Theories with Matter)}},\ }\href
  {https://doi.org/10.1016/0550-3213(85)90316-5} {\bibfield  {journal}
  {\bibinfo  {journal} {Nucl. Phys. B}\ }\textbf {\bibinfo {volume} {260}},\
  \bibinfo {pages} {157} (\bibinfo {year} {1985})}\BibitemShut {NoStop}%
\bibitem [{\citenamefont {Shifman}\ and\ \citenamefont
  {Vainshtein}(1999)}]{Shifman:1999mv}%
  \BibitemOpen
  \bibfield  {author} {\bibinfo {author} {\bibfnamefont {M.~A.}\ \bibnamefont
  {Shifman}}\ and\ \bibinfo {author} {\bibfnamefont {A.~I.}\ \bibnamefont
  {Vainshtein}},\ }\bibfield  {title} {\bibinfo {title} {{Instantons versus
  supersymmetry: Fifteen years later}},\ }in\ \href@noop {} {\emph {\bibinfo
  {booktitle} {{World Sci.Lect.Notes Phys. 62 (1999)}}}}\ (\bibinfo {year}
  {1999})\ pp.\ \bibinfo {pages} {485--647},\ \bibinfo {note} {sections 3.5.4
  and 4.3.},\ \Eprint {https://arxiv.org/abs/hep-th/9902018}
  {arXiv:hep-th/9902018} \BibitemShut {NoStop}%
\bibitem [{\citenamefont {Nekrasov}(2003)}]{Nekrasov:2002qd}%
  \BibitemOpen
  \bibfield  {author} {\bibinfo {author} {\bibfnamefont {N.~A.}\ \bibnamefont
  {Nekrasov}},\ }\bibfield  {title} {\bibinfo {title} {{Seiberg-Witten
  prepotential from instanton counting}},\ }\href
  {https://doi.org/10.4310/ATMP.2003.v7.n5.a4} {\bibfield  {journal} {\bibinfo
  {journal} {Adv. Theor. Math. Phys.}\ }\textbf {\bibinfo {volume} {7}},\
  \bibinfo {pages} {831} (\bibinfo {year} {2003})},\ \Eprint
  {https://arxiv.org/abs/hep-th/0206161} {arXiv:hep-th/0206161} \BibitemShut
  {NoStop}%
\bibitem [{\citenamefont {Shifman}(2022)}]{Shifman:2012zz}%
  \BibitemOpen
  \bibfield  {author} {\bibinfo {author} {\bibfnamefont {M.}~\bibnamefont
  {Shifman}},\ }\href {https://doi.org/10.1017/9781108885911} {\emph {\bibinfo
  {title} {{Advanced topics in quantum field theory}: {A lecture course (2nd
  ed.)}}}}\ (\bibinfo  {publisher} {Cambridge Univ. Press},\ \bibinfo {address}
  {Cambridge, UK},\ \bibinfo {year} {2022})\BibitemShut {NoStop}%
\bibitem [{\citenamefont {Cecotti}\ \emph {et~al.}(1992)\citenamefont
  {Cecotti}, \citenamefont {Fendley}, \citenamefont {Intriligator},\ and\
  \citenamefont {Vafa}}]{Cecotti:1992qh}%
  \BibitemOpen
  \bibfield  {author} {\bibinfo {author} {\bibfnamefont {S.}~\bibnamefont
  {Cecotti}}, \bibinfo {author} {\bibfnamefont {P.}~\bibnamefont {Fendley}},
  \bibinfo {author} {\bibfnamefont {K.~A.}\ \bibnamefont {Intriligator}},\ and\
  \bibinfo {author} {\bibfnamefont {C.}~\bibnamefont {Vafa}},\ }\bibfield
  {title} {\bibinfo {title} {{A New supersymmetric index}},\ }\href
  {https://doi.org/10.1016/0550-3213(92)90572-S} {\bibfield  {journal}
  {\bibinfo  {journal} {Nucl. Phys. B}\ }\textbf {\bibinfo {volume} {386}},\
  \bibinfo {pages} {405} (\bibinfo {year} {1992})},\ \Eprint
  {https://arxiv.org/abs/hep-th/9204102} {arXiv:hep-th/9204102} \BibitemShut
  {NoStop}%
\bibitem [{\citenamefont {Konishi}(1984)}]{Konishi:1983hf}%
  \BibitemOpen
  \bibfield  {author} {\bibinfo {author} {\bibfnamefont {K.}~\bibnamefont
  {Konishi}},\ }\bibfield  {title} {\bibinfo {title} {{Anomalous Supersymmetry
  Transformation of Some Composite Operators in SQCD}},\ }\href
  {https://doi.org/10.1016/0370-2693(84)90311-3} {\bibfield  {journal}
  {\bibinfo  {journal} {Phys. Lett. B}\ }\textbf {\bibinfo {volume} {135}},\
  \bibinfo {pages} {439} (\bibinfo {year} {1984})}\BibitemShut {NoStop}%
\bibitem [{\citenamefont {Seiberg}(1994)}]{Seiberg:1994bz}%
  \BibitemOpen
  \bibfield  {author} {\bibinfo {author} {\bibfnamefont {N.}~\bibnamefont
  {Seiberg}},\ }\bibfield  {title} {\bibinfo {title} {{Exact results on the
  space of vacua of four-dimensional SUSY gauge theories}},\ }\href
  {https://doi.org/10.1103/PhysRevD.49.6857} {\bibfield  {journal} {\bibinfo
  {journal} {Phys. Rev. D}\ }\textbf {\bibinfo {volume} {49}},\ \bibinfo
  {pages} {6857} (\bibinfo {year} {1994})},\ \Eprint
  {https://arxiv.org/abs/hep-th/9402044} {arXiv:hep-th/9402044} \BibitemShut
  {NoStop}%
\bibitem [{\citenamefont {Ritz}\ \emph {et~al.}(2004)\citenamefont {Ritz},
  \citenamefont {Shifman},\ and\ \citenamefont {Vainshtein}}]{Ritz:2004mp}%
  \BibitemOpen
  \bibfield  {author} {\bibinfo {author} {\bibfnamefont {A.}~\bibnamefont
  {Ritz}}, \bibinfo {author} {\bibfnamefont {M.}~\bibnamefont {Shifman}},\ and\
  \bibinfo {author} {\bibfnamefont {A.}~\bibnamefont {Vainshtein}},\ }\bibfield
   {title} {\bibinfo {title} {{Enhanced worldvolume supersymmetry and
  intersecting domain walls in N=1 SQCD}},\ }\href
  {https://doi.org/10.1103/PhysRevD.70.095003} {\bibfield  {journal} {\bibinfo
  {journal} {Phys. Rev. D}\ }\textbf {\bibinfo {volume} {70}},\ \bibinfo
  {pages} {095003} (\bibinfo {year} {2004})},\ \Eprint
  {https://arxiv.org/abs/hep-th/0405175} {arXiv:hep-th/0405175} \BibitemShut
  {NoStop}%
\bibitem [{\citenamefont {Davies}\ \emph {et~al.}(1999)\citenamefont {Davies},
  \citenamefont {Hollowood}, \citenamefont {Khoze},\ and\ \citenamefont
  {Mattis}}]{Davies:1999uw}%
  \BibitemOpen
  \bibfield  {author} {\bibinfo {author} {\bibfnamefont {N.~M.}\ \bibnamefont
  {Davies}}, \bibinfo {author} {\bibfnamefont {T.~J.}\ \bibnamefont
  {Hollowood}}, \bibinfo {author} {\bibfnamefont {V.~V.}\ \bibnamefont
  {Khoze}},\ and\ \bibinfo {author} {\bibfnamefont {M.~P.}\ \bibnamefont
  {Mattis}},\ }\bibfield  {title} {\bibinfo {title} {{Gluino condensate and
  magnetic monopoles in supersymmetric gluodynamics}},\ }\href
  {https://doi.org/10.1016/S0550-3213(99)00434-4} {\bibfield  {journal}
  {\bibinfo  {journal} {Nucl. Phys. B}\ }\textbf {\bibinfo {volume} {559}},\
  \bibinfo {pages} {123} (\bibinfo {year} {1999})},\ \Eprint
  {https://arxiv.org/abs/hep-th/9905015} {arXiv:hep-th/9905015} \BibitemShut
  {NoStop}%
\bibitem [{\citenamefont {Davies}\ \emph {et~al.}(2003)\citenamefont {Davies},
  \citenamefont {Hollowood},\ and\ \citenamefont {Khoze}}]{Davies:2000nw}%
  \BibitemOpen
  \bibfield  {author} {\bibinfo {author} {\bibfnamefont {N.~M.}\ \bibnamefont
  {Davies}}, \bibinfo {author} {\bibfnamefont {T.~J.}\ \bibnamefont
  {Hollowood}},\ and\ \bibinfo {author} {\bibfnamefont {V.~V.}\ \bibnamefont
  {Khoze}},\ }\bibfield  {title} {\bibinfo {title} {{Monopoles, affine algebras
  and the gluino condensate}},\ }\href {https://doi.org/10.1063/1.1586477}
  {\bibfield  {journal} {\bibinfo  {journal} {J. Math. Phys.}\ }\textbf
  {\bibinfo {volume} {44}},\ \bibinfo {pages} {3640} (\bibinfo {year}
  {2003})},\ \Eprint {https://arxiv.org/abs/hep-th/0006011}
  {arXiv:hep-th/0006011} \BibitemShut {NoStop}%
\bibitem [{\citenamefont {Unsal}(2009)}]{Unsal:2007jx}%
  \BibitemOpen
  \bibfield  {author} {\bibinfo {author} {\bibfnamefont {M.}~\bibnamefont
  {Unsal}},\ }\bibfield  {title} {\bibinfo {title} {{Magnetic bion
  condensation: A New mechanism of confinement and mass gap in four
  dimensions}},\ }\href {https://doi.org/10.1103/PhysRevD.80.065001} {\bibfield
   {journal} {\bibinfo  {journal} {Phys. Rev. D}\ }\textbf {\bibinfo {volume}
  {80}},\ \bibinfo {pages} {065001} (\bibinfo {year} {2009})},\ \Eprint
  {https://arxiv.org/abs/0709.3269} {arXiv:0709.3269 [hep-th]} \BibitemShut
  {NoStop}%
\bibitem [{\citenamefont {Cs\'aki}\ \emph {et~al.}(2018)\citenamefont
  {Cs\'aki}, \citenamefont {Martone}, \citenamefont {Shirman},\ and\
  \citenamefont {Terning}}]{Csaki:2017mik}%
  \BibitemOpen
  \bibfield  {author} {\bibinfo {author} {\bibfnamefont {C.}~\bibnamefont
  {Cs\'aki}}, \bibinfo {author} {\bibfnamefont {M.}~\bibnamefont {Martone}},
  \bibinfo {author} {\bibfnamefont {Y.}~\bibnamefont {Shirman}},\ and\ \bibinfo
  {author} {\bibfnamefont {J.}~\bibnamefont {Terning}},\ }\bibfield  {title}
  {\bibinfo {title} {{Pre-ADS Superpotential from Confined Monopoles}},\ }\href
  {https://doi.org/10.1007/JHEP05(2018)188} {\bibfield  {journal} {\bibinfo
  {journal} {JHEP}\ }\textbf {\bibinfo {volume} {05}},\ \bibinfo {pages}
  {188}},\ \Eprint {https://arxiv.org/abs/1711.11048} {arXiv:1711.11048
  [hep-th]} \BibitemShut {NoStop}%
\bibitem [{\citenamefont {Shirman}\ and\ \citenamefont
  {Waterbury}(2019)}]{Shirman:2019mqv}%
  \BibitemOpen
  \bibfield  {author} {\bibinfo {author} {\bibfnamefont {Y.}~\bibnamefont
  {Shirman}}\ and\ \bibinfo {author} {\bibfnamefont {M.}~\bibnamefont
  {Waterbury}},\ }\bibfield  {title} {\bibinfo {title} {{Deformations of the
  moduli space and superpotential flows in 3D SUSY QCD}},\ }\href
  {https://doi.org/10.1103/PhysRevD.99.125002} {\bibfield  {journal} {\bibinfo
  {journal} {Phys. Rev. D}\ }\textbf {\bibinfo {volume} {99}},\ \bibinfo
  {pages} {125002} (\bibinfo {year} {2019})},\ \Eprint
  {https://arxiv.org/abs/1903.11080} {arXiv:1903.11080 [hep-th]} \BibitemShut
  {NoStop}%
\bibitem [{\citenamefont {Bashmakov}\ \emph {et~al.}(2019)\citenamefont
  {Bashmakov}, \citenamefont {Benini}, \citenamefont {Benvenuti},\ and\
  \citenamefont {Bertolini}}]{Bashmakov:2018ghn}%
  \BibitemOpen
  \bibfield  {author} {\bibinfo {author} {\bibfnamefont {V.}~\bibnamefont
  {Bashmakov}}, \bibinfo {author} {\bibfnamefont {F.}~\bibnamefont {Benini}},
  \bibinfo {author} {\bibfnamefont {S.}~\bibnamefont {Benvenuti}},\ and\
  \bibinfo {author} {\bibfnamefont {M.}~\bibnamefont {Bertolini}},\ }\bibfield
  {title} {\bibinfo {title} {{Living on the walls of super-QCD}},\ }\href
  {https://doi.org/10.21468/SciPostPhys.6.4.044} {\bibfield  {journal}
  {\bibinfo  {journal} {SciPost Phys.}\ }\textbf {\bibinfo {volume} {6}},\
  \bibinfo {pages} {044} (\bibinfo {year} {2019})},\ \Eprint
  {https://arxiv.org/abs/1812.04645} {arXiv:1812.04645 [hep-th]} \BibitemShut
  {NoStop}%
\bibitem [{\citenamefont {Dorey}(1998)}]{Dorey:1998yh}%
  \BibitemOpen
  \bibfield  {author} {\bibinfo {author} {\bibfnamefont {N.}~\bibnamefont
  {Dorey}},\ }\bibfield  {title} {\bibinfo {title} {{The BPS spectra of
  two-dimensional supersymmetric gauge theories with twisted mass terms}},\
  }\href {https://doi.org/10.1088/1126-6708/1998/11/005} {\bibfield  {journal}
  {\bibinfo  {journal} {JHEP}\ }\textbf {\bibinfo {volume} {11}},\ \bibinfo
  {pages} {005}},\ \Eprint {https://arxiv.org/abs/hep-th/9806056}
  {arXiv:hep-th/9806056} \BibitemShut {NoStop}%
\bibitem [{\citenamefont {Witten}(1979)}]{Witten:1978bc}%
  \BibitemOpen
  \bibfield  {author} {\bibinfo {author} {\bibfnamefont {E.}~\bibnamefont
  {Witten}},\ }\bibfield  {title} {\bibinfo {title} {{Instantons, the Quark
  Model, and the 1/n Expansion}},\ }\href
  {https://doi.org/10.1016/0550-3213(79)90243-8} {\bibfield  {journal}
  {\bibinfo  {journal} {Nucl. Phys. B}\ }\textbf {\bibinfo {volume} {149}},\
  \bibinfo {pages} {285} (\bibinfo {year} {1979})}\BibitemShut {NoStop}%
\bibitem [{\citenamefont {Hori}\ and\ \citenamefont
  {Vafa}(2000)}]{Hori:2000kt}%
  \BibitemOpen
  \bibfield  {author} {\bibinfo {author} {\bibfnamefont {K.}~\bibnamefont
  {Hori}}\ and\ \bibinfo {author} {\bibfnamefont {C.}~\bibnamefont {Vafa}},\
  }\href@noop {} {\bibinfo {title} {{Mirror symmetry}}} (\bibinfo {year}
  {2000}),\ \Eprint {https://arxiv.org/abs/hep-th/0002222}
  {arXiv:hep-th/0002222} \BibitemShut {NoStop}%
\end{thebibliography}%

\end{document}